\documentclass[9pt,twocolumn,twoside]{osajnl}

\journal{ol} 

\setboolean{shortarticle}{true}
\usepackage{xcolor}
\newcommand{\YS}[1]{\textcolor{black}{ #1}}
\newcommand{\red}[1]{\textcolor{black}{ #1}}

\title{Coherent ray-wave structured light based on (helical) Ince-Gaussian modes}

\author[1,2]{Zhaoyang Wang}
\author[3,4]{Yijie Shen}
\author[1,2]{Qiang Liu}
\author[1,2,5]{Xing Fu}

\affil[1]{Key Laboratory of Photonic Control Technology (Tsinghua University), Ministry of Education, Beijing 100084, China}
\affil[2]{State Key Laboratory of Precision Measurement Technology and Instruments, Department of Precision Instrument, Tsinghua University, Beijing 100084, China}
\affil[3]{Optoelectronics Research Centre, University of Southampton, Southampton SO17 1BJ, UK}

\affil[4]{y.shen@soton.ac.uk}
\affil[5]{fuxing@mail.tsinghua.edu.cn}

\begin{abstract}
\YS{The topological evolution of classic eigenmodes including Hermite-Laguerre-Gaussian and (helical) Ince-Gaussian modes is exploited to construct coherent state modes, which unifies the representations of traveling-wave (TW) and standing-wave (SW) ray-wave structured light for the first time and realizes the TW-SW unified ray-wave geometric beam with topology of ray-trajectories splitting effect, breaking the boundary of TW and SW structured light.} We experimentally generate these new modes with high purity and dynamic control by digital holography method, revealing potential applications in optical manipulation and communication.
\end{abstract}

\setboolean{displaycopyright}{true}

\begin{document}

\maketitle

Structured light, with the ability to arbitrarily tailor its amplitude, phase and polarization~\cite{forbes2020structured}, has attracted great attention due to its wide applications \YS{such as optical tweezers, quantum entanglement, and communications~\cite{shen2019optical,otte2020optical,otte2018entanglement,ndagano2017characterizing}}. Multifarious beam modes as eigensolutions of paraxial wave equation (PWE) were identified in structured light family, such as Hermite-Gaussian (HG) mode, Laguerre-Gaussian (LG) mode, Ince-Gaussian (IG) mode and helical-Ince-Gaussian (HIG) mode~\cite{bandres2004incePWE}, as well as various superposed state of eigenmodes including mixing HG mode~\cite{kotlyar2014hermite}, vortex lattice~\cite{shen2018vortex}, and SU(2) geometric beam~\cite{chen2004wave,shen2018periodic}. The topological evolution and unified representation for prior light modes \YS{rised} enthusiasm. For example, \YS{the generalized Hermite-Laguerre-Gaussian (HLG) modes unify LG and HG modes~\cite{alieva2005mode,abramochkin2017closed}}. IG modes interpret the transition from LG to HG modes based on the topology of elliptical coordinate~\cite{bandres2004incePWE}. A singularity hybrid evolution model, as a further generalized family, accommodates the HLG and HIG modes together~\cite{shen2019hybrid}. Recently, an SU(2) Poincar{\'e} sphere model was proposed to universally reveal the topology of orbital angular momentum (OAM) eigenmodes and coherent state complex modes~\cite{shen20202}. Nevertheless, unified representation for more generalized structured modes still requires further exploration. 

\begin{figure}
\centering
\includegraphics[width=1.0\linewidth]{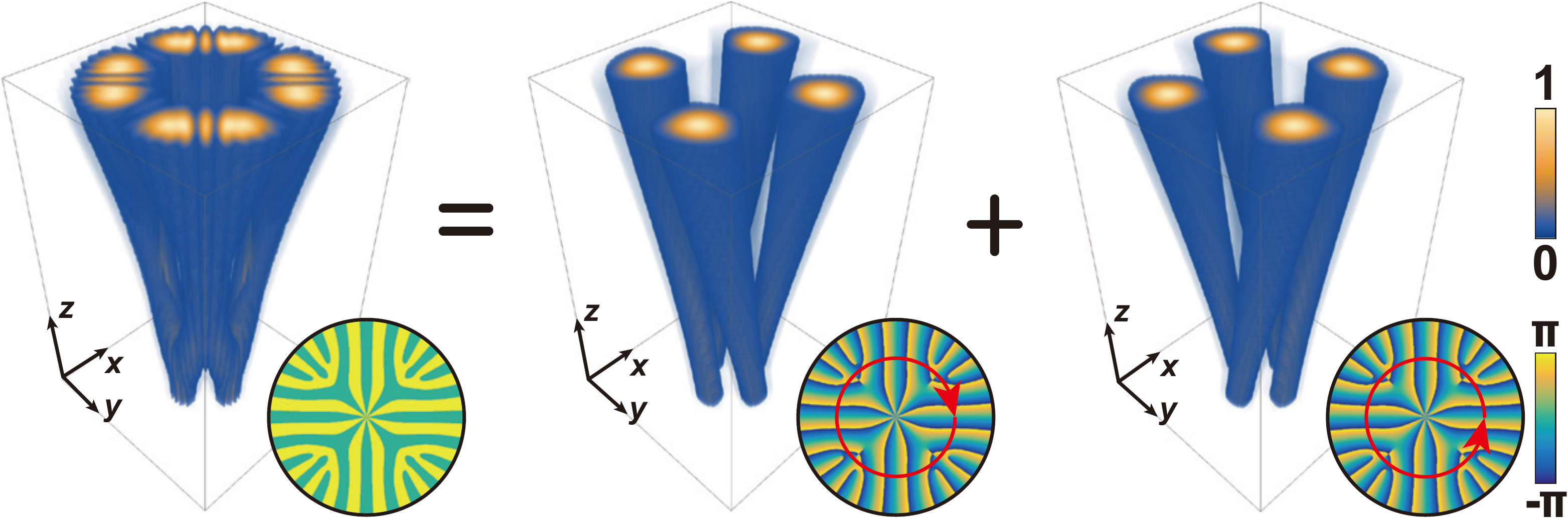}
\caption{\YS{A SW ray-wave mode is composed by two TW ray-wave modes in opposite directions. $z$ ranges from $0$ to $2z_R$. The lower-right inserts show corresponding phases at focus.}}
\label{f.co}
\end{figure}

As another important category of structured light, \YS{the standing-wave (SW) modes, the superposed state of two TW modes propagating at opposite directions, has distinct characteristics and intriguing patterns individually. For instances, the complex SW mode composed by vector vortex TW modes was generated to simulate entangled beating and applied in optical machining~\cite{otte2018entanglement}.} \red{This concept can also be introduced in exotic ray-wave structured light, as shown in Figs.~\ref{f.co}, and the TW and SW ray-wave geometric beams unveil the ray-like structures of light for propagating in freespace~\cite{chen2007generating,shen2018truncated} and oscillating in resonator~\cite{chen2004wave,lu2015generating}, respectively. However, the topological connection of TW and SW ray-wave modes has never been studied in either theory or experiment.} \red{Notably, 
such ray-wave structure has recently extened the frontier of modern physics such as nondiffraction effect~\cite{2020Shaping}, topological phase~\cite{malhotra2018measuring}, and quantum-classical informatics~\cite{shen2020high}. Therefore, breaking the TW-SW boundary in ray-wave structured light is significant for unveiling more general topological evolution and potential applications.}

In this Letter, a unified representation \YS{for topological evolution of TW and SW structured light is proposed and we construct the novel geometric beams with splitting ray-wave structure and mixing ray-wave structure that have not been observed before. We explore the exotic ray-wave structure in generalized geometric beams by constructing the cluster of classical trajectories.} The simulated and experimental results intuitively demonstrate the topological evolution of our generalized structured light.

The PWE has various analytical forms in different coordinates. In cylindrical coordinate $(r,\theta,z)$, \YS{the eigenmodes are TW LG modes and SW LG modes~\cite{beijersbergen1993astigmatic}. The TW LG modes contain spiral phase factor $\text{e}^ {\text{i} \ell \theta}$, noted as $\text{LG}_{p, \ell,l}$ where $p,\ell,l$ are radial, azimuthal, and longitudinal indices ($p = \min(n,m)$ and $\ell = m-n$). The SW LG modes include even and odd LG modes noted as $\text{LG}^{\text{e}}_{p,\ell,l}$ and $\text{LG}^{\text{o}}_{p,\ell,l}$, respectively. $\text{LG}^{\text{e}}_{p,\ell,l}$ contains factor $\cos{(\ell \theta)}$ instead of $\text{e}^ {\text{i}\ell \theta}$ while the $\text{LG}^{\text{o}}_{p,\ell,l}$ contains factor $\sin{(\ell \theta)}$. A SW LG mode is equivalent to the superposition of two TW LG modes with opposite directions. In the elliptic coordinate $(\xi,\eta,z)$ ($\xi \in [0,\infty)$, $\eta \in [0,2\pi)$), which is related to the Cartesian coordinate by elliptic eccentricity $\epsilon$, the eigenmodes are given by IG modes as even IG mode $\text{IG}^{\text{e}}_{u,v,l}(x,y,z|\epsilon)$ and odd IG mode $\text{IG}^{\text{o}}_{u,v,l}(x,y,z|\epsilon)$, where $0\leq v\leq u$ for even IG mode, $1\leq v\leq u$ for odd IG mode, and $(-1)^{u-v}=1$ since $(u,v)$ are of the same parity~\cite{bandres2004incePWE}. Akin to TW and SW LG modes, a mixing mode, HIG mode, can be composed by odd and even IG modes as $\text{HIG}_{u,v,l}^{\pm}=\frac{1}{2}(\text{IG}^{\text{e}}_{u,v,l} \pm \text{i}\text{IG}^{\text{o}}_{u,v,l})$~\cite{bandres2004incePWE}.} In Cartesian coordinate $(x,y,z)$, the eigensolution of PWE is $\text{HG}_{n,m,l}$ mode~\cite{siegman1986lasers}. \YS{In addition, the mixing linear superposition of eigenmodes is also the solution of PWE. More exact topological evolution between these various eigenmodes is still needed to research. Hereinafter, we will study the topological evolution between these classical eigenmodes for a generalized eigenmode family.}

\begin{figure*}
	\centering
	\includegraphics[width=\linewidth]{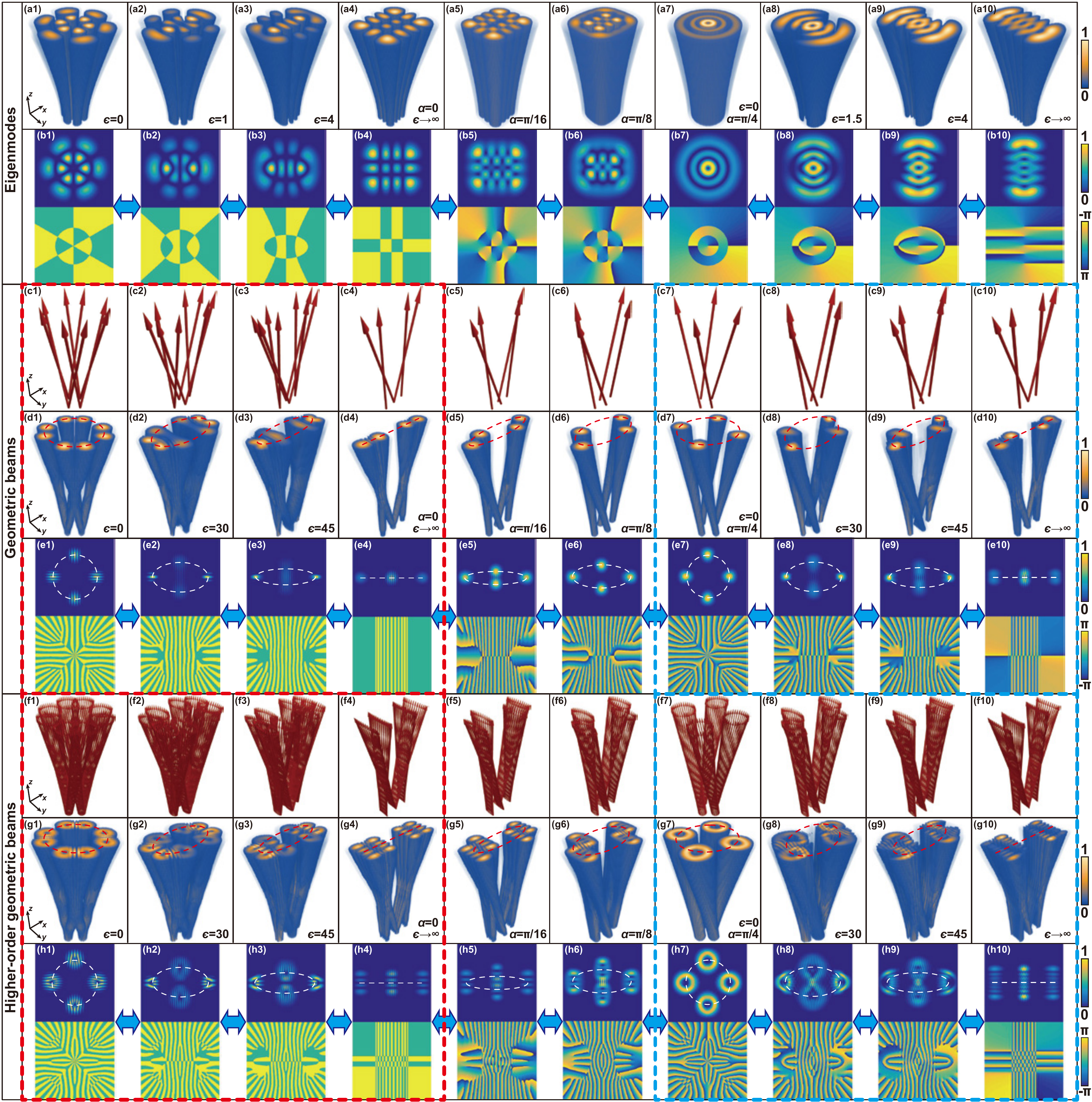}
\caption{\YS{The topological evolutions of TW-SW unified eigenmodes (a, b), geometric beams $(n,m)=(10,0)$ (c, d, e), and higher-order geometric beams $(n,m)=(10,3)$ (f, g, h). Panels (a1-a10), (d1-d10), and (g1-g10) show the corresponding three-dimensional spatial wave packets. Panels (c1-c10) and (f1-f10) show the corresponding classical trajectories for \red{geometric beams and higher-order geoemtric beams}, respectively. $z$ ranges from $0$ to $2z_R$. Panels (b1-b10), \red{(e1-e10), and (h1-h10)} show the corresponding transverse intensity and phase distributions at $z=0$ plane. The subplots in red box exhibit the splitting ray-wave structure and the subplots in blue box exhibit the mixing ray-wave structure. (Colormap: darkness to brightness means 0 to 1 for intensity and $-\pi$ to $\pi$ for phase.)}}
\label{f.theory}
\end{figure*}

\YS{An elliptic coordinate can topologically evolve into Cartesian coordinate for $\epsilon \to \infty$, and to cylindrical coordinate for $\epsilon \to 0$, respectively~\cite{bandres2004incePWE,shen2019hybrid}. Thus IG mode can be considered as the transitional state between TW HG mode and SW LG mode noted as $\psi^{\text{IG}}_{n,m,l}(x,y,z|\epsilon)$ by changing the parameter $\epsilon$, as shown in Figs.~\ref{f.theory} a1-a4 and b1-b4. Besides, HG modes can be converted to TW LG modes via HLG modes~\cite{alieva2005mode,abramochkin2017closed} noted as $\psi^{\text{HLG}}_{n,m,l}(x,y,z|\alpha)$, as shown in Figs.~\ref{f.theory} a4-a7 and b4-b7. Furthermore, the TW LG modes can be further extended to mixing HG mode by exploiting HIG mode noted as $\psi^{\text{HIG}}_{n,m,l}(x,y,z|\epsilon)$, as shown in Figs.~\ref{f.theory} a7-a10 and b7-b10. The generalized eigenmode family $\left\{ \psi^{\text{IG}}_{n,m,l},\psi^{\text{HLG}}_{n,m,l},\psi^{\text{HIG}}_{n,m,l} \right\}$ can be exploited to construct complex coherent state light, further enriching the structured family.}

\begin{figure*}
	\centering
	\includegraphics[width=0.95\linewidth]{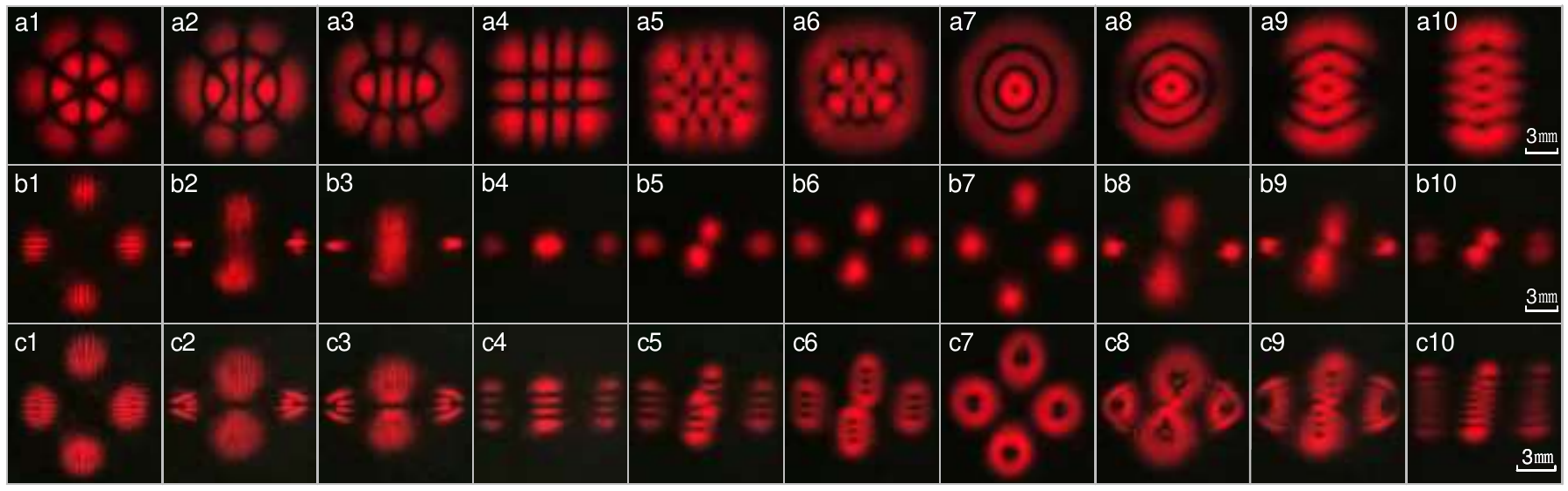}
\caption{Experimental results of the topological evolutions of TW-SW unified eigenmodes (a), geometric beams (b), and higher-order geometric beams (c). See Visualization 1, Visualization 2, Visualization 3 for dynamic movies of (a1-a10), (b1-b10), and (c1-c10).}
\label{f.exp}
\end{figure*}

\textbf{To unify TW and SW ray-wave structured light.} In above sections, we present \YS{the generalized eigenmode family to unify the topological evolution} of TW and SW eigenmodes, which allows us to explore more complex coherent structured mode as on-demand superposed spatial wavepacket of eigenmodes. Hereinafter, we will demonstrate a family of TW-SW unified ray-wave structured geometric beams with striking ray-trajectory splitting and merging properties that have not been observed before. We exploit the formation of SU(2) coherent state to construct such modes~\cite{buvzek1989generalized,chen2004wave,shen20202}:
\begin{equation}
\Phi_{n,m,l}(x,y,z)  =\frac{1}{{{2}^{{N}/{2}\;}}}\sum\limits_{K=0}^{N}{{{\left( \begin{matrix}
			N  \\
			K  \\
			\end{matrix} \right)}^{\frac{1}{2}}}{{\text{e}}^{\text{i}K\phi }} \Psi^{\text{eigen}}_{n+pK,m+qK,l-sK}(x,y,z) },
\label{phi}
\end{equation}
where $\phi$ is coherent phase, $(p,q,s)$ are three integers related to frequency coupling among transverse and longitudinal modes \YS{$Q=p+q$, $s=-P$}, $(P,Q)$ are a pair of coprime integers for fulfilling frequency-degenerate condition~\cite{chen2004wave}, $\Psi^{\text{eigen}}_{n,m,l} \in \left\{ \psi^{\text{IG}}_{n,m,l},\psi^{\text{HLG}}_{n,m,l},\psi^{\text{HIG}}_{n,m,l} \right\}$. The corresponding topological evolution of \YS{spatial wavepacket} for $(n,m)=(10,0)$ \YS{and $(p,q)=(Q,0)$} is shown in Figs.~\ref{f.theory} \YS{d1-d10 and e1-e10.} \YS{The SW and TW geometric beams with oscillating and propagating ray paths were studied in ~\cite{chen2007generating,shen2018truncated,lu2015generating}. But we create new modes as the transitional state between them}, as shown in Figs.~\ref{f.theory} c1-c4 \YS{and d1-d4}, ray orbits of a planar TW geometric beam gradually split into two fold and distribute into oscillating trajectory of the circular SW geometric beam, \red{which unify the topological evolution of TW and SW geometric beams.} Besides, the generalized geometric beam also includes the evolution from a planar trajectory beam into the circular trajectory beam carrying OAM (Figs.~\ref{f.theory} c4-c7). \red{Furthermore, we could utilize HIG modes to extend more exotic} transformation from the OAM state into a mixing planar trajectory beam (Figs.~\ref{f.theory} c7-c10), not proposed before either.

We also explore the higher-order formation of such coherent state mode, as the results for $(n,m)=(10,3)$ \YS{and $(p,q)=(Q,0)$} demonstrated in Figs.~\ref{f.theory} \YS{f1-f10, g1-g10 and h1-h10}, where the light on each ray state changes into higher-order HLG mode formation, namely the multi-axis vortex beam~\cite{tuan2018characterization}. Here we markedly generalize such multi-vortex geometric beam that the TW multi-HG beam can topologically evolve into SW multi-LG beam (Figs.~\ref{f.theory} g1-g4) and can also evolve into TW multi-LG beam (Figs.~\ref{f.theory} g4-g7) and further into multi-mixing HG beam (Figs.~\ref{f.theory} g7-g10). \YS{The newly proposed ray-wave geometric beams (in red and blue boxes of Figs.~\ref{f.theory}) largely enrich the structured light family and inspire the tailoring of more exotic structured light.}


\textbf{To explore ray-wave structure with splitting ray orbits.}
\YS{Due to the quantum-classical correspondence nature of coherent state, the coherent state geometric beam shows intriguing ray-wave duality~\cite{chen2004wave}, i.e. the wave pattern of which is localized on a cluster of classical ray trajectories. For TW modes, the cluster of classical trajectories $\left\{ x^{\text{b}}_s,y^{\text{b}}_s,z|\alpha \right\}^{\pm}$ is coupled with spatial wave packet of Eq.~(\ref{phi}) for $\Psi^{\text{eigen}}_{n,m,l} = \psi^{\text{HLG}}_{n,m,l}(x,y,z|\alpha)$ (geometric beams based on HLG modes)~\cite{chen2019laser}, where $\alpha$ ranges from $-\pi/4$ to $\pi/4$, $\pm$ represents two opposite directions, \red{$(x^{\text{b}}_s,y^{\text{b}}_s,z)$ represents a classical trajectory labelled $s$ where spatial wave packets are located on, and $\left\{ x^{\text{b}}_s,y^{\text{b}}_s,z|\alpha \right\}^{\pm}$ a collection of such classical trajectories, where $s=0,1,\cdots,Q-1$ labels the ray number (see detailed expressions in Supplement).} Here we select $\alpha$ ranging from $0$ to $\pi/4$ for TW modes, corresponding to the mode evolution from planar to circular classical orbits, as shown in Figs.~\ref{f.theory} c4-c7 and f4-f7. Furthermore, we can construct the cluster of classical trajectories for generalized geometric beams.}

\YS{SW modes can be decomposed into two TW modes, which reveals that the cluster of classical trajectories of SW geometric beams (shown in Figs.~\ref{f.theory} c1 and f1) can be interpreted as a superposition of two clusters of classical trajectories of TW modes in opposite directions as:
\begin{equation}
\left\{ x^{\text{b}}_s,y^{\text{b}}_s,z \right\}^{\text{SW}}=\left\{ x^{\text{b}}_s,y^{\text{b}}_s,z|\alpha=\pi/4 \right\}+\left\{ x^{\text{b}}_s,y^{\text{b}}_s,z|\alpha=-\pi/4 \right\}.
\label{cla_SW}
\end{equation}
Since SW modes correspond to the case $\epsilon=0$, the $\left\{ x^{\text{b}}_s,y^{\text{b}}_s,z \right\}^{\text{SW}}$ can be noted as a limiting case $\left\{ x^{\text{b}}_s,y^{\text{b}}_s,z|\epsilon=0 \right\}^{\text{IG}}$, where $\left\{ x^{\text{b}}_s,y^{\text{b}}_s,z|\epsilon \right\}^{\text{IG}}$ is the cluster of classical trajectories coupled with spatial wave packet of Eq.~(\ref{phi}) for $\Psi^{\text{eigen}}_{n,m,l} = \psi^{\text{IG}}_{n,m,l}(x,y,z|\epsilon)$ (geometric beams based on IG modes), revealing the exotic splitting orbits in evolution of SW-TW modes with $\epsilon$ increasing from $0$ to $\infty$ \red{(see details in Supplement)}, as shown in Figs.~\ref{f.theory} c1-c4, f1-f4.} 

\red{Besides, the mixing HG modes (HIG modes with $\epsilon \to \infty$) are essentially equal to a superposition of two HG modes with different indices, as shown in Figs.~\ref{f.theory} c8-c10, f8-f10, where the cluster of classical trajectories coupled with spatial wave packet of Eq.~(\ref{phi}) for $\Psi^{\text{eigen}}_{n,m,l} = \psi^{\text{HIG}}_{n,m,l}(x,y,z|\epsilon)$ (geometric beams based on HIG modes), revealing exotic mixing orbits with $\epsilon$ increasing from $0$ to $\infty$. The exotic TW-SW-unified ray-wave structures reveal the generalized ray-wave duality in generalized geometric beam, providing a deeper physical insight of quantum-classical correspondence (ray-wave duality).}

\textbf{Experimental realization.} We experimentally generate these complex modes with high-purity based on the classic digital holography method by a digital micromirror device~\cite{ren2015tailoring,wan2020}. Experimental results of TW-SW unified structured light are shown in Figs.~\ref{f.exp}, where rows from top to bottom are the patterns of \YS{TW-SW} unified eigenmodes, multi-path geometric beams and multi-HLG higher-order geometric beams, respectively, recorded at $z=0$ plane, corresponding to simulated results in Figs.~\ref{f.theory}.

\red{\textbf{Discussion.} Our model of TW-SW-unified structured light is largely extendable. For instance, it can also be applied to more complex SU(2) coherent state corresponding to generalized ray-wave Lissajous and trochoidal wavepacket~\cite{chen2010spatial,wan2020digitally}. We can also study its general structure in astigmatic and vectorial optical fields. In addition, other kinds of coherent superposed formations are also expected to explore, such as SU(1,1) coherent state~\cite{wodkiewicz1985coherent}, and hybrid coherent state~\cite{shen2020structured}. The TW-SW unification also act as a new mechanism to extend topological structure, so as to enable novel applications. The ray-trajectory-splitting topology provides new degrees of freedom to create multi-partite classical entangled state~\cite{shen2020high}, which can be employed in high-speed optical encryption and communication~\cite{ndagano2017characterizing}. The multi-singularity and complex OAM evolution of the new structured light is also in need of advanced optical tweezers and trapping~\cite{otte2020optical}}.


In summary, we propose a new theory to unify the TW and SW formations of structured light. \YS{It generalizes the family of ray-wave geometric modes based on TW-SW-unified eigenmodes (IG, HLG, and HIG modes), extending the new topological ray-wave structures as thier complex coherent states.} The generalized theoretical framework has strong extensibility and applicability to construct more complex modes and to study OAM with multi-singularities, which inspires the exploration of more topological properties of novel structured modes with their advanced applications.

\textbf{Funding.}
The National Key Research and Development Program of China (2017YFB1104500); 
National Natural Science Foundation of China (61975087); 
Beijing Young Talents Support Project (2017000020124G044).

\noindent\textbf{Disclosures.} The authors declare no conflicts of interest.

\bibliography{sample}

\begin{thebibliography}{10}
\newcommand{\enquote}[1]{``#1''}

\bibitem{forbes2020structured}
A.~Forbes, {\protect\JournalTitle{Optics and Photonics News}} \textbf{31}, 24
  (2020).

\bibitem{shen2019optical}
Y.~Shen, X.~Wang, Z.~Xie, C.~Min, X.~Fu, Q.~Liu, M.~Gong, and X.~Yuan,
  {\protect\JournalTitle{Light: Science \& Applications}} \textbf{8}, 90
  (2019).

\bibitem{otte2020optical}
E.~Otte and C.~Denz, {\protect\JournalTitle{Applied Physics Reviews}}
  \textbf{7}, 041308 (2020).

\bibitem{otte2018entanglement}
E.~Otte, C.~Rosales-Guzm{\'a}n, B.~Ndagano, C.~Denz, and A.~Forbes,
  {\protect\JournalTitle{Light: Science \& Applications}} \textbf{7}, 18009
  (2018).

\bibitem{ndagano2017characterizing}
B.~Ndagano, B.~Perez-Garcia, F.~S. Roux, M.~McLaren, C.~Rosales-Guzman,
  Y.~Zhang, O.~Mouane, R.~I. Hernandez-Aranda, T.~Konrad, and A.~Forbes,
  {\protect\JournalTitle{Nature Physics}} \textbf{13}, 397 (2017).

\bibitem{bandres2004incePWE}
M.~A. Bandres and J.~C. Guti{\'e}rrez-Vega, {\protect\JournalTitle{JOSA A}}
  \textbf{21}, 873 (2004).

\bibitem{kotlyar2014hermite}
V.~Kotlyar and A.~Kovalev, {\protect\JournalTitle{JOSA A}} \textbf{31}, 274
  (2014).

\bibitem{shen2018vortex}
Y.~Shen, Z.~Wan, X.~Fu, Q.~Liu, and M.~Gong, {\protect\JournalTitle{JOSA B}}
  \textbf{35}, 2940 (2018).

\bibitem{chen2004wave}
Y.-F. Chen, C.~Jiang, Y.-P. Lan, and K.-F. Huang,
  {\protect\JournalTitle{Physical Review A}} \textbf{69}, 053807 (2004).

\bibitem{shen2018periodic}
Y.~Shen, X.~Yang, X.~Fu, and M.~Gong, {\protect\JournalTitle{Applied Optics}}
  \textbf{57}, 9543 (2018).

\bibitem{alieva2005mode}
T.~Alieva and M.~J. Bastiaans, {\protect\JournalTitle{Optics Letters}}
  \textbf{30}, 1461 (2005).

\bibitem{abramochkin2017closed}
E.~Abramochkin and T.~Alieva, {\protect\JournalTitle{Optics Letters}}
  \textbf{42}, 4032 (2017).

\bibitem{shen2019hybrid}
Y.~Shen, Y.~Meng, X.~Fu, and M.~Gong, {\protect\JournalTitle{JOSA A}}
  \textbf{36}, 578 (2019).

\bibitem{shen20202}
Y.~Shen, Z.~Wang, X.~Fu, D.~Naidoo, and A.~Forbes,
  {\protect\JournalTitle{Physical Review A}} \textbf{102}, 031501 (2020).

\bibitem{chen2007generating}
C.-H. Chen and C.-F. Chiu, {\protect\JournalTitle{Optics Express}} \textbf{15},
  12692 (2007).

\bibitem{shen2018truncated}
Y.~Shen, X.~Fu, and M.~Gong, {\protect\JournalTitle{Optics Express}}
  \textbf{26}, 25545 (2018).

\bibitem{lu2015generating}
T.-H. Lu and C.~He, {\protect\JournalTitle{Optics Express}} \textbf{23}, 20876
  (2015).

\bibitem{2020Shaping}
A.~Zannotti, C.~Denz, M.~A. Alonso, and M.~R. Dennis,
  {\protect\JournalTitle{Nature Communications}} \textbf{11} (2020).

\bibitem{malhotra2018measuring}
T.~Malhotra, R.~Guti{\'e}rrez-Cuevas, J.~Hassett, M.~Dennis, A.~Vamivakas, and
  M.~Alonso, {\protect\JournalTitle{Physical review letters}} \textbf{120},
  233602 (2018).

\bibitem{shen2020high}
Y.~Shen, I.~Nape, X.~Yang, X.~Fu, M.~Gong, D.~Naidoo, and A.~Forbes,
  {\protect\JournalTitle{Light Science \& Applications, (in press)}}  (2021).

\bibitem{beijersbergen1993astigmatic}
M.~W. Beijersbergen, L.~Allen, H.~Van~der Veen, and J.~Woerdman,
  {\protect\JournalTitle{Optics Communications}} \textbf{96}, 123 (1993).

\bibitem{siegman1986lasers}
A.~E. Siegman, \emph{Lasers} (University Science, Mill Valley, CA, 1986).

\bibitem{buvzek1989generalized}
V.~Bu{\v{z}}ek and T.~Quang, {\protect\JournalTitle{JOSA B}} \textbf{6}, 2447
  (1989).

\bibitem{tuan2018characterization}
P.~Tuan, Y.~Hsieh, Y.~Lai, K.-F. Huang, and Y.-F. Chen,
  {\protect\JournalTitle{Optics Express}} \textbf{26}, 20481 (2018).

\bibitem{chen2019laser}
Y.~Chen, S.~Li, Y.~Hsieh, J.~Tung, H.~Liang, and K.-F. Huang,
  {\protect\JournalTitle{Optics Letters}} \textbf{44}, 2649 (2019).

\bibitem{ren2015tailoring}
Y.-X. Ren, R.-D. Lu, and L.~Gong, {\protect\JournalTitle{Annalen Der Physik}}
  \textbf{527}, 447 (2015).

\bibitem{wan2020}
Z.~Wan, Z.~Wang, X.~Yang, Y.~Shen, and X.~Fu, {\protect\JournalTitle{Optics
  Express}}  (2020).

\bibitem{chen2010spatial}
Y.-F. Chen, Y.~Lin, K.-F. Huang, and T.-H. Lu, {\protect\JournalTitle{Physical
  Review A}} \textbf{82}, 043801 (2010).

\bibitem{wan2020digitally}
Z.~Wan, Z.~Wang, X.~Yang, Y.~Shen, and X.~Fu, {\protect\JournalTitle{Optics
  Express}} \textbf{28}, 31043 (2020).

\bibitem{wodkiewicz1985coherent}
K.~Wodkiewicz and J.~Eberly, {\protect\JournalTitle{JOSA B}} \textbf{2}, 458
  (1985).

\bibitem{shen2020structured}
Y.~Shen, X.~Yang, D.~Naidoo, X.~Fu, and A.~Forbes,
  {\protect\JournalTitle{Optica}} \textbf{7}, 820 (2020).

\end{thebibliography}

\bibliographyfullrefs{sample}

\end{document}